# The transport-structural correspondence across the nematic phase transition probed by elasto-x-ray diffraction

Joshua J Sanchez[1], Paul Malinowski[1], Joshua Mutch[1], Jian Liu[2], J-W. Kim[3], Philip J Ryan[3,4], Jiun-Haw Chu[1,*]

**Affiliation:**

[1] Department of Physics, University of Washington, Seattle, Washington 98195, USA.

[2] Department of Physics and Astronomy, University of Tennessee, Knoxville, Tennessee 37996, USA.

[3] Advanced Photon Source, Argonne National Laboratories, Lemont, Illinois 60439, USA.

[4] School of Physical Sciences, Dublin City University, Dublin 9, Ireland.

*Correspondence to: jhchu@uw.edu (J.-H.C)

**Abstract: (148/150 words)**

Electronic nematicity in iron pnictide materials is coupled to both the lattice and the conducting electrons, which allows both structural and transport observables to probe nematic fluctuations and the order parameter. Here we combine simultaneous transport and x-ray diffraction measurements with in-situ tunable strain (elasto-XRD) to measure the temperature dependence of the shear modulus and elastoresistivity above the nematic transition and the spontaneous orthorhombicity and resistivity anisotropy below the nematic transition, all within a single sample of $\mathrm{Ba(Fe_{0.96}Co_{0.04})_2As_2}$. The ratio of transport to structural quantities is nearly temperature-independent over a 74 K range and agrees between the ordered and disordered phases. These results show that elasto-XRD is a powerful technique to probe the nemato-elastic and nemato-transport couplings, which have important implications to the nearby superconductivity. It also enables the measurement in the large strain limit, where the breakdown of mean field description reveals the intertwined nature of nematicity.

## Main Text: (2897/3000 words)

Electronic nematicity refers to a spontaneous rotational symmetry breaking phase in solids driven by electronic correlations[1]. It was initially discovered in a few fine-tuned systems, such as the quantum Hall states of the two-dimensional electron gas[2,3] and the field induced metamagnetic state in $Sr_3Ru_2O_7$[4], but its existence was later observed in the broader class of strongly correlated systems, including iron-based[5,6] and copper-based[7,8] high temperature superconductors and more recently in magic angle twisted bi-layer graphene[9]. In almost all cases, the first signature of nematicity has been the onset of large resistivity anisotropy, which is regarded as a proxy of the nematic order parameter. The correspondence between the nematicity and transport anisotropy originates from symmetry considerations. The nematic order parameter $\psi$ and the resistivity anisotropy $\eta = \frac{\rho_{xx}-\rho_{yy}}{\rho_{xx}+\rho_{yy}}$ belong to the same irreducible representation of the high symmetry point group, hence they are linearly proportional to each other in the infinitesimal limit[10] ($\eta = k\psi$). Nevertheless, the resistivity is not a thermodynamic variable and depends on extrinsic properties such as disorder[11–14]. Therefore, a key question is to what extent a transport coefficient can represent the order parameter beyond the infinitesimal limit. This question is especially important for 2D systems such as graphene where thermodynamic measurements are difficult.

Here we examine this question in a model system for the study of nematic phase transitions, the iron pnictide superconductor[15,16]. Because of the electron-lattice coupling, the nematicity in iron-pnictides has been clearly observed in both transport anisotropy and structural thermodynamic variables (Fig.4a). Above the phase transition, the diverging nematic susceptibility can be seen in the Curie-Weiss temperature dependence of both the $2m_{66}$ elastoresistivity coefficient[17] and the softening of the shear modulus[18] $C_{66}$. Below the phase transition, the nematic order parameter also generates a large spontaneous resistivity anisotropy[6,19] $\eta_S$ and a spontaneous structural distortion[20,21] $\varepsilon_S$. Each of these four quantities are usually measured separately due to incompatible sample preparation needed for standard techniques, which makes a quantitative comparison difficult. In particular, the spontaneous resistivity anisotropy $\eta_S$ is notoriously difficult to measure close to the phase transition, because the stress required to detwin the sample always induces additional resistivity anisotropy due to the softening of $C_{66}$ and the divergence of $2m_{66}$. To our knowledge, no study of iron pnictides has ever reported any two of the above quantities within a single sample. In this work, we perform measurements of $2m_{66}$, $C_{66}$, $\eta_S$ and $\varepsilon_S$ using the technique of elasto x-ray diffraction all within one single crystal sample of $\text{Ba}(\text{Fe}_{0.96}\text{Co}_{0.04})_2\text{As}_2$, located on the underdoped side of the phase diagram with $T_S = 73.8 \text{ K}$, $T_N = 64 \text{ K}$ and $T_C = 13 \text{ K}$ (Fig. 1a). With our unprecedented multi-modal measurement, we show the four quantities perfectly follows a mean-field temperature dependence. Furthermore, the ratio of transport to structural quantities is a constant across the phase transition, suggesting that the resistivity anisotropy behaves just like a thermodynamic variable even for large values of the nematic order parameter. While the $2m_{66}$, $C_{66}$, $\eta_S$ and $\varepsilon_S$ can be well described by the Landau free energy framework, two unexpected findings stand out. First, using the $C_{66}$ and $2m_{66}$ data from the previous studies[17,22], we discovered a strong doping dependence of the ratio between transport and structural quantities, increasing by more than fivefold towards optimal doping. Second, when driving the system deep into the non-linear regime with large uniaxial stress, we found that the resistivity anisotropy shows a non-saturating behavior that is drastically different from the dampened response of the lattice. Possible implications of these two unusual phenomena are discussed.

**Elasto X-Ray Diffraction**

In order to simultaneously assess the electronic and structural response to stress in a single crystal sample, we have developed a new experimental platform, the elasto X-ray diffraction (elasto-XRD), that combines electrical transport with *in-situ* uniaxial stress tunability via a Razorbill CS-100 strain device[23] fully integrated with x-ray diffraction (XRD) measurements at beamline 6-ID-B at the Advanced Photon Source (Fig. 1b) (see Methods). This platform allows us to measure the lattice constants, orthorhombic twin domain populations, and electrical resistivity simultaneously while the uniaxial stress is continuously tuned to detwin the sample and enhance the orthorhombicity. At fixed temperatures from 66 K to 140 K, XRD and transport measurements were made by ramping strain from maximum compression to maximum tension and back. Uniaxial stress was applied along the tetragonal $[1\ 1\ 0]_T$ direction ($\hat{x}$). The nominal strain is defined as $\varepsilon_{xx}^{nom} = \frac{\Delta L}{L_0}$, where $L_0$ is the size of the gap between two titanium plates on which the sample was glued with Stycast epoxy. The displacement $\Delta L$ was determined from a capacitance strain gauge. The four-wire electrical contact geometry enables the simultaneous resistance measurements along the stress axis.

**Nematic fluctuation divergence: shear modulus and elastoresistivity**

We focus first on the structural and electronic response to applied stress above the nematic transition, where there are no orthorhombic domains. We define $\varepsilon_{xx} = \frac{\Delta a}{a_0}$ and $\varepsilon_{yy} = \frac{\Delta b}{b_0}$ as the XRD-measured strains of the inline ($[1\ 1\ 0]_T$) and transverse ($[-1\ 1\ 0]_T$) lattice constants, respectively. At 130 K the lattice constants show a nearly linear response to $\varepsilon_{xx}^{nom}$, while just above the transition at 74 K the response becomes strongly nonlinear, with an enhanced response at $\varepsilon_{xx}^{nom} = 0$ (Fig.2a). This results in the strain transmission $\frac{d\varepsilon_{xx}}{d\varepsilon_{xx}^{nom}}$ and the induced $B_{2g}$ orthorhombicity $\varepsilon = \frac{1}{2}(\varepsilon_{xx} - \varepsilon_{yy})$ becoming increasingly nonlinear with cooling (Supplementary Figure 2). The rate of change of the in-plane transverse strain to inline strain $\frac{d\varepsilon_{yy}}{d\varepsilon_{xx}}$ approaches a peak value of -1 with cooling to $T_S$ and decreases to its high-temperature value at large $\varepsilon$ (Fig. 2b).

We extract the shear modulus from Poisson's ratio ($\nu_{xy} = -\frac{d\varepsilon_{yy}}{d\varepsilon_{xx}}|_{\varepsilon=0}$) using $C_{66} = (50.5\ GPa)\frac{1-\nu_{xy}}{1+\nu_{xy}}$, where the magnitude is determined from other elastic modulus terms using ultrasound data from ref.[22] (see Supplementary Information Section VII). In Figure 2c, $C_{66}$ diminishes to nearly zero at $T_S$ (black dots). Fitting $C_{66}$ with a Curie-Weiss temperature dependence, $C_{66} = C_{66,0} - A(T - T^*)^{-1}$ (red line, $R^2 > .98$),

yields a fitted value of the bare shear modulus $C_{66,0} = 38.8 \pm 4.7$ GPa in agreement with the high-temperature ultrasound data[22]. The extracted bare nematic transition temperature, $T^* = 50K \pm 8.3K$ is considerably larger than the values obtained from several other shear modulus measurements[5,18,24,25], where $T_S - T^* \sim 40K - 50K$, yet as we discuss below it agrees well with the $T^*$ obtained from the simultaneous elastoresistance measurement. We note that a major difference is that in the previously reported measurements $C_{66}$ is reduced but nonzero at the transition, possibly due to local strain inhomogeneities and resulting domain microstructures adding a small background signal near the transition[25]. Since $C_{66}$ varies most rapidly near the transition, this difference may strongly influence the Curie-Weiss fitting, hence the discrepancy in $T^*$.

We next turn to the resistivity response to strain. The resistivity $\rho_{xx}$ vs $\varepsilon_{xx}^{nom}$ is increasingly nonlinear with cooling (Fig.2d). In particular, near the phase transition $\rho_{xx}$ shows a kink-like behavior as $\varepsilon_{xx}^{nom}$ increases from zero to positive, and an inflection point at large negative values of $\varepsilon_{xx}^{nom}$. Intriguingly, while still being nonlinear, the kink and inflection point of $\rho_{xx}$ vanish when plotted against the simultaneously measured orthorhombicity $\varepsilon$ (Fig.2e), and $\rho_{xx}$ can be well fitted by a 2$^{nd}$ order polynomial (Supplementary Figure 3). This stark contrast indicates that the anomalies observed in $\rho_{xx}$ vs $\varepsilon_{xx}^{nom}$ are artifacts of the nonlinear strain transmission, highlighting the importance of in-situ x-ray measurements. The 2$^{nd}$ order polynomial can be understood from a symmetry analysis by decomposing the inline resistivity dependence on the B$_{2g}$ orthorhombicity as

$$\rho_{xx}(\varepsilon) = \rho_0 \left(1 + m_{B_{2g}}^{B_{2g}}\varepsilon + m_{A_{1g}}^{B_{2g},B_{2g}}\varepsilon^2\right)$$

where $m_{B_{2g}}^{B_{2g}} = 2m_{66}$ is the linear coefficient with (odd) B$_{2g}$ symmetry and $m_{A_{1g}}^{B_{2g},B_{2g}}$ is the quadratic coefficient with (even) A$_{1g}$ symmetry. Consistent with previous work[26], $m_{A_{1g}}^{B_{2g},B_{2g}}$ has a large magnitude near the transition. As we cannot perform a simultaneous bidirectional transport measurement with this setup, we isolate the B$_{2g}$ component and extract the $2m_{66}$ elastoresistivity coefficient at each temperature using $2m_{66} = \frac{d}{d\varepsilon}\left(\frac{\Delta\rho_{xx}}{\rho_0}\right)|_{\varepsilon=0}$ , which diverges toward the transition (Fig.2f). We analyze this temperature dependence with the same Curie-Weiss model as used for the shear modulus and find it is well described with a similar $T^* = 48.9K \pm 7.1K$ (Fig.3e, blue line, $R^2 > .95$). Therefore, we demonstrate that both 2m$_{66}$ and C$_{66}$ within a single sample show the same mean-field temperature dependence, confirming both have a linear proportionality to a common driver, namely the nematic fluctuations.

While the 2m$_{66}$ and C$_{66}$ extracted near the zero-strain limit show good agreement with mean field behavior, it is no longer the case in the large strain limit where nematic fluctuations are expected to be heavily dampened. The Figure 2e inset shows the resistivity anisotropy $\eta = \frac{\rho_{xx}(\varepsilon)-\rho_{xx}(-\varepsilon)}{\rho_{xx}(\varepsilon)+\rho_{xx}(-\varepsilon)}$ , defined as the normalized resistivity difference at equal in-plane orthorhombicity between tension and compression, and its strain derivative $\frac{d\eta}{d\varepsilon}$ which corresponds to the induced nematic order parameter and nematic susceptibility at finite strain (note that $\frac{d\eta}{d\varepsilon}|_{\varepsilon=0} = \frac{d}{d\varepsilon}\left(\frac{\Delta\rho_{xx}}{\rho_0}\right)|_{\varepsilon=0} = 2m_{66}$). At temperatures near or below $T_S$, $\eta$ shows no sign of saturation as $\varepsilon$ exceeds 0.18% (the spontaneous orthorhombicity at T = 8 K, which is also the maximum value of spontaneous orthorhombicity for this doping concentration). This non-saturating behavior is in sharp contrast to the structural counterpart, where $\frac{d\varepsilon_{yy}}{d\varepsilon_{xx}}$ dampens rapidly towards its high temperature value (Fig.2b). Although in the large-stress limit $\frac{d\varepsilon_{yy}}{d\varepsilon_{xx}}$ and $\frac{d\eta}{d\varepsilon}$ no longer simply relate to $C_{66}$ and $2m_{66}$, which are response functions defined in the zero-stress limit, the striking difference between $\frac{d\varepsilon_{yy}}{d\varepsilon_{xx}}$ and $\frac{d\eta}{d\varepsilon}$ is unexpected. This peculiar finding will be revisited in the discussion section.

**Spontaneous Elastoresistivity**

Next, we extract the spontaneous orthorhombicity $\varepsilon_S$ and spontaneous resistivity anisotropy $\eta_S$ in the nematic ordered phase. We focus on the 10 K range below $T_S$ but above $T_N$ because the long-range antiferromagnetic order induces shifts in the orthorhombicity[20,21] and reconstructs the Fermi surface leading to additional resistivity anisotropy effects [27–30]. Upon cooling the sample below $T_S = 73.8K$, the single peak of the $(2\ 2\ 12)_T$ reflection splits into two peaks corresponding to the $a_A$ and $b_B$ orthorhombic lattice constants of the A and B domains, respectively, indicating the formation of structural twin domains (Fig. 3a). As is often done for a freestanding crystal[21], we define the spontaneous orthorhombicity as $\varepsilon_S = \frac{a_A - b_B}{a_A + b_B}$ in the zero-stress limit. Within the purely nematic phase, $\varepsilon_S$ is well fitted to a mean-field $\sqrt{T_S - T}$ temperature dependence (Fig. 3f).

The presence of twin domains cause transport measurements to average over the resistivities along the two domain directions. This presents a substantial experimental challenge to obtaining the resistivity anisotropy of the orthorhombic unit cell, which we overcome by precisely strain detwinning the sample. Figure 3c-e shows detwinning results for a representative temperature (66 K). The peak positions and intensities ($I_A$ and $I_B$) are shown in Figure 3c. Strain homogeneity is confirmed by a nearly constant Bragg peak width throughout the nominal strain range (Fig. 3d). The relative volume fraction of the A domain is determined as $D_A = \left(\frac{I_A}{I_A + I_B}\right) \times 100\%$, which varies smoothly between 0% and 100% with applied stress, i.e. between the B and A monodomains (Fig. 3e, right). While the sample is mostly detwinned over a relatively small strain range, the last 10% volume fraction of the minor domain is detwinned over larger strain values where the inline lattice constant becomes highly susceptible to applied strain. Therefore, we find that we can mostly, but not fully, detwin the sample without inducing additional lattice distortions, which warrants consideration for the design and interpretation of future experiments involving uniaxial stress detwinning of structural domains.

The spontaneous resistivity anisotropy $\eta_S = \frac{\rho_a - \rho_b}{\rho_a + \rho_b}$ results from the different resistivities along the $\hat{a}$ and $\hat{b}$ directions, $\rho_a$ and $\rho_b$. Due to the network of twin domains running at 45° to the length of the sample[31], the current takes nontrivial paths which results in a nonlinear dependence of $\rho_{xx}$ on $D_A$ (Supplementary Figure 7a). We approach the problem from two directions. First, we start in a zero nominal strain state and detwin the sample just until the lattice begins to deform (at 85% and 93% full detwinning to the compressive B and tensile A domains, respectively; see Supplementary Figure 6). This is the closest condition to a single domain state without lattice distortion that we can achieve experimentally. The extracted $\rho_a$ and $\rho_b$ (Fig. 3b) yield values of $\eta_S$ which are well fit to $\sqrt{T_S - T}$ (Fig.3g). Alternatively, we can extract $\eta_S$ by starting in the fully-detwinned regime and linear fitting the resistivity anisotropy $\eta$ down to $\varepsilon_S$. The resulting values of $\eta_S$ are also well fit by $\sqrt{T_S - T}$ (Supplementary Figure 7b-b). The $\eta_S$ obtained by these two approaches agree within 5%, suggesting that the remnant minor domain has a minimal impact on the transport. This result stands in sharp contrast to earlier works using fixed-strain/stress detwinning in a clamp or horseshoe device[6,11,19,32–35], which generally found a large resistivity anisotropy above $T_S$ from the strain-induced $2m_{66}$ that mixes with the detwinned domain $\eta_S$ below $T_S$, preventing a determination of the real mean-field development of $\eta_S$. Thus, our elasto-XRD technique allows for a precise measurement of a new transport coefficient, the *spontaneous elastoresistivity*, defined as a resistivity anisotropy $\eta_S$ and structural order parameter $\varepsilon_S$ driven by the system itself in the zero-stress limit. In the next section we describe the physical interpretation of the spontaneous elastoresistivity and how it is related to $2m_{66}$ and $C_{66}$ above the transition.

## Temperature independent transport-structural proportionality

The shared mean-field temperature dependence of $\eta_S$ and $\varepsilon_S$ below the transition echo the shared Curie-Weiss temperature dependence of the $2m_{66}$ and $C_{66}$ above the transition, demonstrating the one-to-one correspondence between the transport and structural coefficients. The remaining question is whether this one-to-one correspondence is continuous across the phase transition. In Figure 4b we plot $\frac{\eta_S}{\varepsilon_S}$, the spontaneous elastoresistivity, and $2m_{66}\left(1 - \frac{C_{66}}{C_{66,0}}\right)^{-1}$. These two quantities are the ratios between the dimensionless transport and structural coefficients below and above $T_S$, respectively. We find that both quantities show almost no temperature dependence and their temperature-averaged values are in strong agreement, with $\frac{\eta_S}{\varepsilon_S} = 142.6 \pm 20.7$ and $2m_{66}\left(1 - \frac{C_{66}}{C_{66,0}}\right)^{-1} = 142.9 \pm 29.7$. This agreement suggests that for the entire temperature range of this study, the resistivity anisotropy behaves like a thermodynamic order parameter for all practical purposes. We note that this relationship is valid even at the lowest measured temperature within the pure nematic phase, 66 K, where the orthorhombicity reaches ~40% of its saturation value at base temperature, well beyond the infinitesimal limit.

We now show that $\frac{\eta_S}{\varepsilon_S}$ and $2m_{66}\left(1 - \frac{C_{66}}{C_{66,0}}\right)^{-1}$ can be taken as the relative coupling of nematicity to the conducting electrons compared to the lattice. We consider the Landau free energy that describes a nematic phase transition with a bilinear coupling to the lattice (full derivation in Methods):

$$F = \frac{a_0(T - T^*)}{2}\psi^2 + \frac{b}{4}\psi^4 + \frac{C_{66,0}}{2}\varepsilon^2 - \lambda\psi\varepsilon.$$

Minimizing the free energy below the phase transition, we obtain the primary nematic order parameter $\psi_S \propto \sqrt{T_S - T}$ which induces a secondary order parameter, the spontaneous orthorhombicity $\varepsilon_S = \frac{\lambda}{C_{66,0}}\psi_S$. Combined with the linear nemato-transport relation $\eta_S = k\psi_S$, we get $\frac{\eta_S}{\varepsilon_S} = \frac{k C_{66,0}}{\lambda}$. Minimizing the free energy above the phase transition yields a Curie-Weiss nematic susceptibility $\frac{d\psi}{d\varepsilon} = \frac{\lambda}{a(T-T^*)}$ which results in $2m_{66} = k\frac{\lambda}{a(T-T^*)}$ and $\frac{C_{66}}{C_{66,0}} = 1 - \frac{\lambda}{C_{66,0}}\frac{\lambda}{a(T-T^*)}$. From this we find $2m_{66}\left(1 - \frac{C_{66}}{C_{66,0}}\right)^{-1} = \frac{k C_{66,0}}{\lambda}$, identical to $\frac{\eta_S}{\varepsilon_S}$ below the transition (Fig. 4a). If we ignore $C_{66,0}$, which is a material specific parameter not related to nematicity, we have a simple physical interpretation of $\frac{\eta_S}{\varepsilon_S}$ and $2m_{66}\left(1 - \frac{C_{66}}{C_{66,0}}\right)^{-1}$ – they measure the ratio of the nemato-transport coupling constant $k$ and nemato-elastic coupling constant $\lambda$.

**Discussion**

Extensive measurements of $2m_{66}$ and $C_{66}$ have been made across the phase diagram of Ba(Fe$_{1-x}$Co$_x$)$_2$As$_2$. In Figure 4c we plot the fitted Curie constant ($k\frac{\lambda}{a}$) for the $2m_{66}$ data from ref.[17] and ($\frac{\lambda}{C_{66,0}}\frac{\lambda}{a}$) for the $C_{66}$ data from ref.[24], which demonstrates that while ($k\frac{\lambda}{a}$) is enhanced towards the optimal doping, ($\frac{\lambda}{C_{66,0}}\frac{\lambda}{a}$) is not similarly enhanced. This observation suggests that the enhancement of $2m_{66}$ with doping is due not only to an enhancement of nematic fluctuations themselves but also due to a relative enhancement of nematic coupling to conduction electrons over the lattice. Indeed, we find that $2m_{66}\left(1-\frac{C_{66}}{C_{66,0}}\right)^{-1}$, or $\frac{k\,C_{66,0}}{\lambda}$, increases by more than a factor of 5 towards the optimal doping. Given that $C_{66,0}$ only shows weak doping dependence[22], the increase of $\frac{k\,C_{66,0}}{\lambda}$ can only come from the relative increase of $k$ over $\lambda$, a condition which favors the superconducting pairing by nematic fluctuations[36]. A similar conclusion was made in a recent work comparing the doping evolution of the elastocaloric effect and elastoresistivity in this same system[37], which found evidence for a diminishing value of $\lambda$ with doping towards optimal. Further, elastoresistivity measurements have shown a similar increase of $k\frac{\lambda}{a}$ near quantum critical points across systems as diverse as $\text{La}(\text{Fe}_{1-x}\text{Co}_x)\text{AsO}$ and $\text{FeSe}_{1-x}\text{S}_x$, suggesting that this enhancement of nematic-transport coupling near the quantum critical point may be quite general in the iron-based superconductors[38,39].

The efficacy of resistivity anisotropy as a representation of a thermodynamic order parameter and its breakdown in the large stress limit has a profound implication to the microscopic mechanism of nematicity. In the framework of Boltzmann transport theory, resistivity anisotropy is determined by the anisotropy of elastic and inelastic scattering rates and Fermi surfaces. Several theoretical studies argued that anisotropic spin fluctuations, the leading candidate of the microscopic mechanism of nematicity in iron pnictides, generate anisotropy in both elastic and inelastic scattering[12,13,40]. This picture provides a natural explanation for the non-saturating resistivity anisotropy in the large stress limit. The large stress shifts the antiferromagnetic transition to a higher temperature, which increases the spin fluctuations[41] and hence induces additional resistivity anisotropy. This is a non-linear effect that arises from the intertwined nature of vestigial nematicity[42], which is not captured in the Landau free energy discussed above. We note that this highly non-linear nemato-elastic coupling has also been observed in a recent elasto-scanning tunneling microscopy measurement.[43] Further, a recent study of FeSe shows a similar breakdown of transport-structural correspondence as order parameter grows beyond 50%, which may be related to the unusual spin fluctuations in this system[44]. Future study using elasto-XRD on multiple material systems will help clarify this issue.

From the experimental perspective, the use of x-ray diffraction gives unprecedented detail in the detwinning process itself and reveals a highly non-linear structural and electronic responses close to the phase transition. While similar uniaxial stress approaches have been used recently to explore interesting properties in iron pnictides and beyond [41,45–52], this work highlights the importance of in-situ microscopic measurement of structurally-complex quantum materials.

## Methods

**Sample Preparation**

Single crystal samples of $Ba(Fe_{.96}Co_{.04})_2As_2$ were grown from an FeAs flux as described elsewhere[17]. The primary sample used in x-ray measurements was prepared as a thin bar of dimensions 2.0 x 0.57 x 0.07 mm and cut along the Fe-Fe bonding direction. Gold wires were glued with DuPont 4929 silver epoxy underneath the sample to not obstruct the x-ray diffraction off the top surface of the crystal. Measurements of the resistivity coefficient $\rho_{xx}$ aligned along the stress axis were performed using a standard 4-point measurement and an SR830 lock-in amplifier.

**X-ray Diffraction**

X-ray diffraction (XRD) measurements were performed at the Advanced Photon Source, beamline 6-ID-B, at Argonne National Laboratories. X-rays of energy 11.215 keV illuminated an area 500x500 um, fully encompassing a cross section of the middle of the crystal where strain transmission is highest. The sample and strain device were mounted on a closed cycle cryostat. Gaussian fits to the tetragonal $(2\ 2\ 12)_T$, $(-1\ 1\ 14)_T$ and $(0\ 0\ 14)_T$ reflections were used to determine the orthorhombic lattice constants in the direction of applied stress ($a_A$ & $b_B$), in-plane transverse to the stress ($a_B$ & $b_A$) and normal to the plane ($c$), corresponding to the $\hat{x}$, $\hat{y}$, and $\hat{z}$ directions, respectively.

**Fit Parameters**

Below are the fitting parameters used in Fig. 2c, 2f, 3f and 3g respectively:

$C_{66} = C_{66,0} + A\left(\frac{1}{T-T^*}\right)$ :$T^* = 50.0K \pm 8.3K$, $A = -933 \pm 51$ GPaK, $C_{66,0} = 38.8 \pm 4.7$ GPa, $R^2 = 0.98$

$2m_{66} = 2m_{66,0} + A\left(\frac{1}{T-T^*}\right)$ : $T^* = 48.9K \pm 7.1K$, $A = 4237 \pm 330$, $2m_{66,0} = -14.3 \pm 8.5$, $R^2 = 0.95$

$\varepsilon_S = A\sqrt{73.8\ K - T}$ , $A = .000290 \pm .0000006$ , $R^2 = 0.99$

$\eta_S = A\sqrt{73.8\ K - T}$ , $A = .0421 \pm .0022$ , $R^2 = 0.97$

The $2m_{66}$ data from ref.[17] for 4.7% Co-doping was reevaluate to yield fit values:

$2m_{66} = 2m_{66,0} + A\left(\frac{1}{T-T^*}\right)$ : $T^* = 36.4K \pm 0.9K$, $A = 4150 \pm 135$, $2m_{66,0} = -23.1 \pm 1.0$, $R^2 = 0.99$

**Domain Detwinning Video**

A video of the detwinning process is available online. At 66 K, strain is applied through a loop from maximum compression to maximum tension and back. Top plot: The log-scale intensity of the split $(2\ 2\ 12)_T$ peak across the whole area detector ("chi" vs $2\theta$). The shifts in intensity across the "chi" direction (y axis) indicate small reorientations of crystal grains, which are summed at each value of $2\theta$ to obtain the total intensity used in the middle plot and the main text Figure 3. Middle plot: Gaussian fits to the chi-summed intensity vs $2\theta$. Bottom plot: Relative amplitudes of the Gaussian fits yields the relative population of the A domain ($D_A = \frac{I_A}{I_A+I_B}x100\%$) vs the measured nominal strain.

**Free Energy Derivation.**

Here we discuss in more detail the Landau free energy described in the main text.

$$F = \frac{a\ (T-T^*)}{2}\psi^2 + \frac{b}{4}\psi^4 + \frac{C_{66,0}}{2}\varepsilon^2 - \lambda\psi\varepsilon - h\varepsilon.$$

In the high-symmetry (tetragonal) phase, there is no static nematic order ($\langle\psi\rangle = 0$). An applied stress $h$ causes the orthorhombicity to become nonzero, which creates nematic order. Minimizing the free energy first with respect to $\psi$ and then to $\varepsilon$ and taking the $\psi = 0$ limit yields the zero-stress nematic susceptibility, $\frac{d\psi}{d\varepsilon} = \frac{\lambda}{a}\left(\frac{1}{T-T^*}\right)$. Taking the resistivity anisotropy $\eta = \frac{\rho_{xx}-\rho_{yy}}{\rho_{xx}+\rho_{yy}}$ to be linearly proportional to the nematic order parameter, $\eta = k\psi$, the elastoresistivity coefficient $2m_{66}$ becomes linearly proportional to the nematic susceptibility, $2m_{66} = \frac{d\eta}{d\varepsilon} = k\frac{d\psi}{d\varepsilon}$. To obtain the renormalized shear modulus $C_{66}$ we minimize the free energy with respect to $\varepsilon$ and then to $\psi$ and use $\frac{dh}{d\psi} = \frac{d\varepsilon}{d\psi}\frac{dh}{d\varepsilon} = \frac{C_{66}}{\left(\frac{d\psi}{d\varepsilon}\right)}$ to obtain

$$C_{66} = C_{66,0} - \lambda\frac{d\psi}{d\varepsilon}.$$

Below the nematic transition under zero stress ($h = 0$), the nematic order parameter becomes spontaneously nonzero ($\psi_S$) which drives the spontaneous orthorhombicity ($\varepsilon_S$). Minimizing the free energy with respect to $\varepsilon$ yields the magnitude of the orthorhombicity, $\varepsilon_S = \frac{\lambda}{C_{66,0}}\psi_S$. Assuming the spontaneous resistivity anisotropy remains proportional as well ($\eta_S = k\psi_S$), the ratio $\frac{\eta_S}{\varepsilon_S} = \frac{k}{\left(\frac{\lambda}{C_{66,0}}\right)}$ becomes temperature-independent and independent of the nematic order parameter. Subbing $\varepsilon = \frac{\lambda}{C_{66,0}}\psi$ into the free energy and minimizing with respect to $\psi$, we obtain

$$\psi\left[a\left(T - T^* - \frac{\lambda^2}{aC_{66,0}}\right) + b\psi^2\right] = 0$$

As the second term in the brackets is always positive (since $b > 0$ is required for stability), we find $\psi$ may only have a nonzero value for $T < T^* + \frac{\lambda^2}{aC_{66,0}} = T_S$, yielding an enhanced nematic transition temperature $T_S$. The nematic order parameter grows with a mean-field temperature dependence as $\psi_S = \sqrt{\frac{a}{b}(T_S - T)}$, resulting in mean-field temperature dependencies of $\varepsilon_S$ and $\eta_S$.

**Acknowledgments:** We thank Cenke Xu, Jing-Yuan Chen, Rafael Fernandes, Anton V. Andreev, and Matthias Ikeda for discussion. **Funding:** This work was mainly supported by NSF MRSEC at UW (DMR-1719797) and the Air Force Office of Scientific Research under Grant FA9550-17-1-0217 and Grant FA9550-21-1-0068. J.H.C. acknowledge the support of the Gordon and Betty Moore Foundation's EPiQS Initiative, Grant GBMF6759 to J.-H.C, the David and Lucile Packard Foundation, the Alfred P. Sloan foundation and the State of Washington funded Clean Energy Institute. J. L. acknowledges support from the National Science Foundation under Grant No. DMR-1848269. This research used resources of the Advanced Photon Source, a U.S. Department of Energy (DOE) Office of Science User Facility operated for the DOE Office of Science by Argonne National Laboratory under Contract No. DE-AC02-06CH11357. J.J.S. was partially supported by the U.S. Department of Energy, Office of Science, Office of Workforce Development for Teachers and Scientists, Office of Science Graduate Student Research (SCGSR) program, administered by the Oak Ridge Institute for Science and Education (ORISE) for the DOE. ORISE is managed by ORAU under contract number DE-SC0014664.

**Author contributions:** J.M. grew the samples. J.J.S. and P.M. did the experiments. P.R., J.-W.K., and J.L. helped conceive and design the XRD measurements at the APS. J.J.S. analyzed the data. J.H.C. supervised the project. All authors contributed extensively to the interpretation of the data and the writing of the manuscript.

**Competing interests:** Authors declare no competing interests.

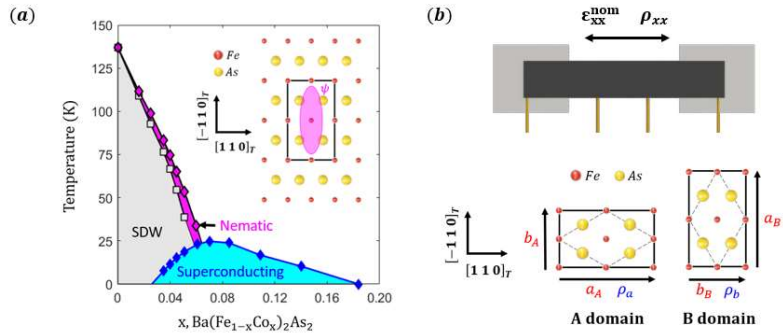

**Figure 1|: Nematic-elastic-transport coupling.** (a) *x-T* phase diagram of Co-doped BaFe$_2$As$_2$. Inset is a representation of the nematic order parameter $\psi$ (magenta) aligned with the in-plane orthorhombicity $\varepsilon$ (black). (b) Schematic of the sample measurement geometry and strain device. Uniaxial stress is applied along the tetragonal $[1\,1\,0]_T$ direction. Inline resistivity $\rho_{xx}$ measures $\rho_a$ of the A domain and $\rho_b$ of the B domain.

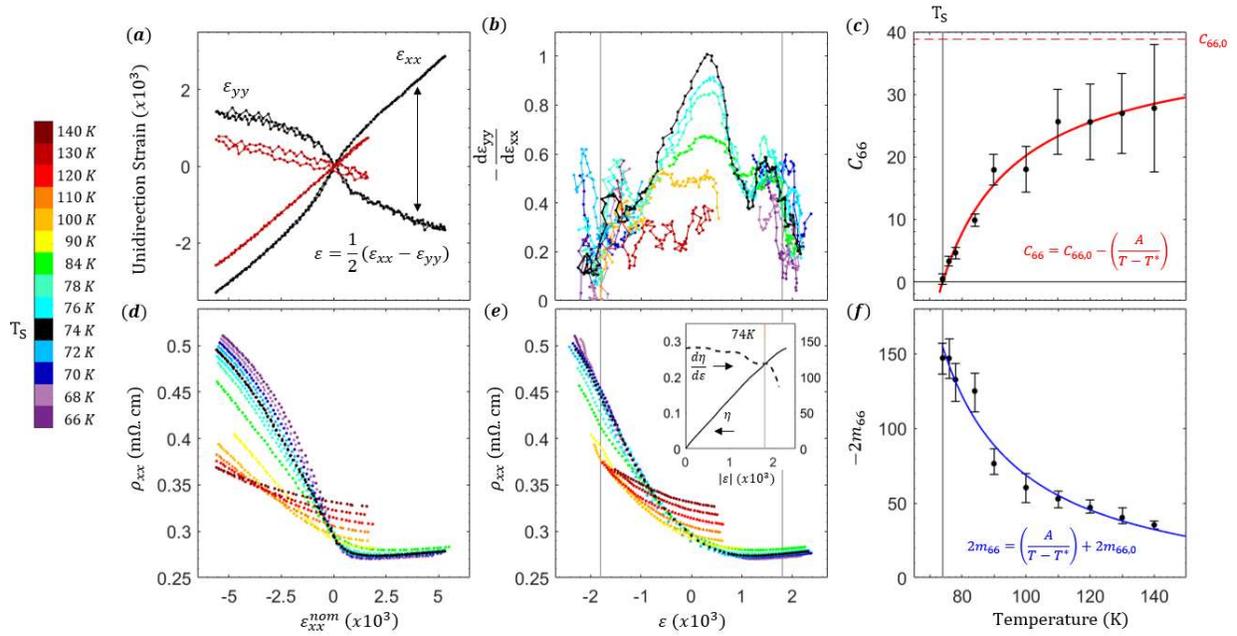

**Figure 2| Shear modulus and elastoresistivity.** (a) Unidirectional lattice constant strains $\varepsilon_{xx}$ and $\varepsilon_{yy}$ vs nominal strain $\varepsilon_{xx}^{nom}$ at $T = 74\ K$ (black) and 130 K (red). In-plane orthorhombicity $\varepsilon = \frac{1}{2}(\varepsilon_{xx} - \varepsilon_{yy})$. (b) The rate of change of in-plane unidirectional strains $-\frac{d\varepsilon_{yy}}{d\varepsilon_{xx}}$ vs $\varepsilon$. (c) The shear modulus $C_{66}$ extracted from $-\frac{d\varepsilon_{yy}}{d\varepsilon_{xx}}|_{\varepsilon=0}$. (d) Longitudinal resistivity $\rho_{xx}$ vs $\varepsilon_{xx}^{nom}$ and (e) vs $\varepsilon$. Inset to (e) shows the resistivity anisotropy $\eta = \frac{\rho_{xx}(\varepsilon) - \rho_{xx}(-\varepsilon)}{\rho_{xx}(\varepsilon) + \rho_{xx}(-\varepsilon)}$ and the derivative $\frac{d\eta}{d\varepsilon}$ at $T = 74$ K. (f) The $2m_{66}$ elastoresistivity extracted from $\frac{d\rho/\rho_0}{d\varepsilon}|_{\varepsilon=0}$. Fit lines in (c) and (f) described in main text. Grey lines in (b) and (e) show the 8 K zero-stress value of in-plane orthorhombicity, $\varepsilon = 0.18\%$.

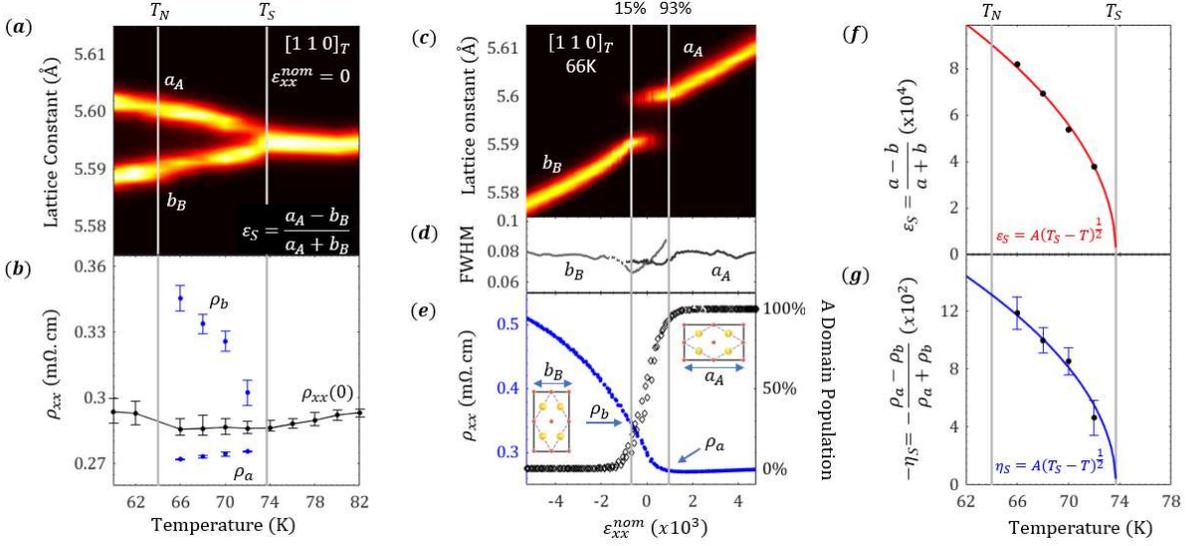

**Figure 3| Spontaneous resistivity anisotropy and orthorhombicity.** (a) Zero-stress x-ray diffraction of the $[1\,1\,0]_T$ lattice constants across the nematic ($T_S$) and antiferromagnetic ($T_N$) transitions. (b) The zero-stress twinned state resistivity $\rho_{xx}$ (black) and the detwinned monodomain resistivities $\rho_a$ and $\rho_b$ (blue) extracted from the $D_A = 90\%$ and $10\%$ points in (e). (c) X-ray diffraction of the $[1\,1\,0]_T$ lattice constants $a_A$ and $b_B$ with intensities $I_A$ and $I_B$. (d) The full-width half maximum (FWHM) of the Gaussian fit to the XRD peak for both $a_A$ and $b_B$. (e, right) Relative A domain population, $D_A = \frac{I_A}{I_A+I_B}$. Grey bars at $D_A = 15\%$ and $D_A = 93\%$. (e, left) Inline resistivity with monodomain resistivities $\rho_a$ and $\rho_b$ at $D_A = 93\%$ and $D_A = 15\%$. (f) The spontaneous orthorhombicity $\varepsilon_S$ and (g) the spontaneous resistivity anisotropy $\eta_S$ are both well fit to a $\sqrt{T_S - T}$ temperature dependence within the nematic phase with $T_S = 73.8K$.

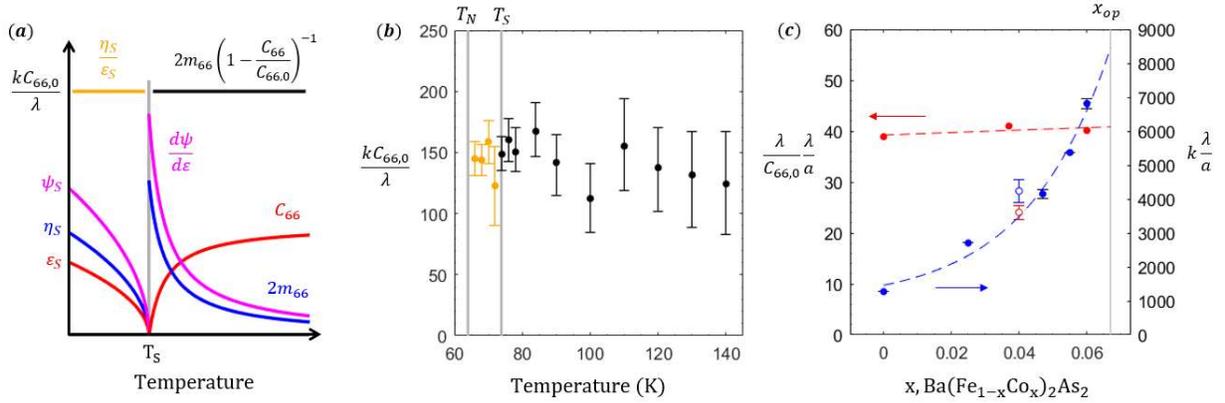

**Figure 4| Transport-structural ratio equivalence.** (a) For $T < T_S$, the spontaneous nematic order parameter $\psi_S \propto \sqrt{T_S - T}$ drives the linear proportional spontaneous orthorhombicity $\varepsilon_S = \frac{\lambda}{C_{66,0}}\psi_S$ and spontaneous resistivity anisotropy $\eta_S = k\,\psi_S$, yielding a temperature independent ratio $\frac{\eta_S}{\varepsilon_S}$. For $T > T_S$ the diverging nematic susceptibility $\frac{d\psi}{d\epsilon} = \frac{\lambda}{a(T^*-T)}$ drives the diverging elastoresistivity $2m_{66} = k\frac{d\psi}{d\epsilon}$ and the softened shear modulus $C_{66} = C_{66,0} - \lambda\frac{d\psi}{d\epsilon}$. The ratio $2m_{66}\left(1 - \frac{C_{66}}{C_{66,0}}\right)^{-1}$ is thus also temperature independent. If the nematic-elastic ($\lambda$) and nematic-transport ($k$) proportionality coefficients are constant across the phase transition, both ratios equate at $T_S$ with a value $\frac{kC_{66,0}}{\lambda}$. (b) The measured ratios $\frac{\eta_S}{\varepsilon_S}$ (gold) and $2m_{66}\left(1 - \frac{C_{66}}{C_{66,0}}\right)^{-1}$ (black) vs temperature. (c) The Curie constants from Curie-Weiss fits to $2m_{66}$ (blue) and $C_{66}$ (red) across the underdoped side of the phase diagram. Data from refs. [17,22] (see main text). Dashed lines are a guide to the eye, solid vertical line at optimal Co doping $x_{op} = 0.067 \pm .02$. Open symbols from this work.

24